\documentclass[aps,prl,twocolumn,showpacs]{revtex4}
\usepackage{graphicx}
\usepackage{dcolumn}
\usepackage{bm}

\include{psfig}
\input epsf

\begin{document}
%
%
%
\title{Physics of the liquid-liquid critical point}

\author{Francesco Sciortino, Emilia La Nave and Piero Tartaglia}

\affiliation{
Dipartimento di Fisica and INFM Udr and 
Center for Statistical Mechanics and Complexity,
Universita' di Roma                     
"La Sapienza" \\
Piazzale Aldo Moro 2, I-00185, Roma, Italy
}

\begin{abstract}
Within the inherent structure ($IS$) thermodynamic formalism introduced by Stillinger and Weber [F. H. Stillinger and T. A. Weber, Phys. Rev. A {\bf 25}, 978 (1982)] we address the basic question of the 
physics of the liquid-liquid transition and of density
maxima observed in some complex liquids such as water 
by identifying,  for the first time, the statistical properties of the potential energy 
landscape (PEL) responsible  for these anomalies. 
 We also provide evidence of the connection between density anomalies and  
the liquid-liquid critical point. Within the simple
(and physically transparent) model  discussed, density anomalies do 
imply the existence of a liquid-liquid transition.

 \end{abstract}
\maketitle

Water, the most important liquid for life, belongs to a class of liquids  for which the isobaric temperature dependence of the density has a maximum. 
In contrast to ordinary liquids, water expands on
cooling below $4 C$ at atmospheric pressure.
This density anomaly is associated with 
other thermodynamic anomalies, such as minima in
the compressibility along isobars and an increase of the specific heat on cooling~\cite{angellreview}.
Recent  fascinating studies have attempted to connect these anomalies to the presence of  two different 
liquid structures, separated at low temperature by
a line of first order transitions, ending in a second order 
critical point~\cite{poole,ponyatonski,mix}. 
In the case of water, this novel critical point would be located in an experimentally  inaccessible region~\cite{mishimastanley}.  Despite this unfavorable location, the postulated presence of this critical point 
provides a framework for interpreting~\cite{poolepre} not only features  of the liquid state but also the phenomenon of polyamorphism and the first-order like  transition between polyamorphs~\cite{mishima}. 
In this Letter we aim at identifying  the statistical properties of the potential energy landscape (PEL) responsible  for   the density maxima and the connection to the  physics of the liquid-liquid transition.
 
The study of the statistical properties of the PEL ---
i.e. of the number, shape and depth of the
basins composing the potential energy hypersurface ---
is under tremendous development. Theoretical approaches, 
based on the seminal work of Stillinger and Weber~\cite{sw}, combined either  with calorimetric data~\cite{speedy,stillingerjpc,lewis} or
with analysis of extensive numerical
simulations~\cite{sastrydeben,ivan,mossa}  are starting to provide precise estimates of the statistical properties of the PEL in several materials and models for liquids. 
A simple model  for the statistical properties of the landscape,  supported by recent numerical studies~\cite{sastrynature,mossa,starr01}, 
can be built  on the basis of the two following hypotheses:
\begin{enumerate}

\item a gaussian distribution for $\Omega(e_{IS})de_{IS}$~\cite{rem,sasai,heuer00,sastrynature,keyes},
the number of distinct basins of energy depth 
$e_{IS}$ between $e_{IS}$ and  $e_{IS}+de_{IS}$, i.e.

\begin{equation}
\Omega(e_{IS}) de_{IS}=e^{\alpha N}
\frac
{e^{-(e_{IS}-E_0)^2/2\sigma^2}}
{\sqrt{2 \pi \sigma^2}}
de_{IS}
\label{eq:Omega}
\end{equation}
Here $e^{\alpha N}$ counts the total number of PEL basins,
($N$ being the number of molecules)
$E_0$ is the characteristic energy scale and $\sigma^2$ is the variance of the distribution. A gaussian distribution is suggested by the central limit theorem, since  in the absence of diverging correlation,  $e_{IS}$ can be written as the sum of the IS energy of several independent subsystems. 
%
\item A (multi dimensional)  parabolic shape of the PEL basins. This hypothesis is equivalent to assuming that the $T$-dependence of basin exploration is governed by an harmonic vibrational  free energy $f_{vib}$. For a system of rigid water molecules
\begin{equation}
f_{vib}(e_{IS},T) =k_B T \sum_{i=1}^{6N-3} 
ln(\beta \hbar \omega_{i}(e_{IS}))]
\label{eq:fvib}
\end{equation}
where $\hbar$ is Planck's constant, $\beta=1/(k_B T)$ and  $k_B$ is the Boltzmann constant and $\omega_i$ is the frequency of the $i$-th normal mode.  As suggested by numerical evidence~\cite{sastrynature,mossa,starr01},  the relation between the parabolic basin  volume and the basin depth is assumed to be linear and the linearity is quantified by two parameters $a$ and $b $  by writing  $\sum_{i=1}^{6N-3} \ln(\omega_i(e_{IS}))  = a + b e_{IS}$.  More accurate descriptions for $f_{vib}(e_{IS},T)$, incorporating anharmonic corrections, can be employed\cite{emipisa} if accuracy in the model is required. In this work,
which focuses on the connection between landscape features and thermodynamic scenarios, the simple
harmonic expression is considered. Accounting for
anharmonic corrections does not affect the main conclusion of this work.
\end{enumerate}

For this harmonic gaussian landscape model, 
the three parameters $\alpha$, $E_0$, $\sigma^2$ 
(modeling the statistical properties of the PEL) and
the two parameters $a$ and  $b$ (modeling the
relation between basin shape and depth) fully determine
the free energy of the liquid.  
The corresponding equation of state (EOS), expressed in term of the $V$-derivative of the landscape properties, is~\cite{eos}

\begin{equation}
P(T,V)= {\cal P}_{const}+T {\cal P}_{_{_{T}}}+T^{-1} {\cal P}_{_{_{1/T}}},
\label{eq:componenti}
\end{equation}
where ${\cal P}_{const}(V)=-  \frac {d} { d V} [E_0-{b\sigma^2}]$, 
$ {\cal P}_{_{_{T}}}(V)=R \frac {d} { d V}  [\alpha -a-{  bE_0+ b^2\sigma^2/2} ]$ and ${\cal P}_{_{_{1/T}}}(V) = \frac {d} { d V} [\sigma^2/2 R]$.  

Along an isochore, the high $T$ behavior is
fixed by the linear term $T {\cal P}_{_{_{T}}}$ , while the
low $T$ behavior is controlled by the  $T^{-1} $ term 
$T^{-1}  {\cal P}_{_{_{1/T}}}$.   From 
Eq.~\ref{eq:componenti} one can also conclude that,
in the harmonic gaussian landscape, $P$ along
an isochore either is monotonically increasing with $T$ 
(if ${\cal P}_{_{_{1/T}}} \le 0$, simple liquid cases), or it has a minimum at  $T=\sqrt{{\cal P}_{_{_{1/T}}}(V)/{\cal P}_{_{_{T}}}(V)}$
(if ${\cal P}_{_{_{1/T}}}>0$, liquids with density anomalies)\cite{nota1}.    

Eq.~\ref{eq:componenti}  offers the possibility of  understanding the landscape parameters  responsible for  density anomalies.  Indeed,  a Maxwell relation states that a
density maximum state point (i.e. a point where $ \partial V/ \partial T |_P=0$)  is simultaneously a point at which
the $T$-dependence of $P$ 
along  an isochore has a minimum  ($\partial P/\partial T|_V=0$).   Hence, the condition for the existence of
density maxima,  from Eq.~\ref{eq:componenti}, is that $P_{_{_{1/T}}} >0$ which, in the PEL formalism, corresponds to $d \sigma^2/dV>0$.
Thus, in normal liquids, in the $V$-range where liquid states exist, $d \sigma^2/dV$ must always be negative.
On the contrary,  liquids with density anomalies must be characterized by  a $V$-range  where  $\sigma^2$ increases with $V$ \cite{nota2}. 
 
\begin{figure}[t] 
\centering
\includegraphics[width=.4\textwidth]{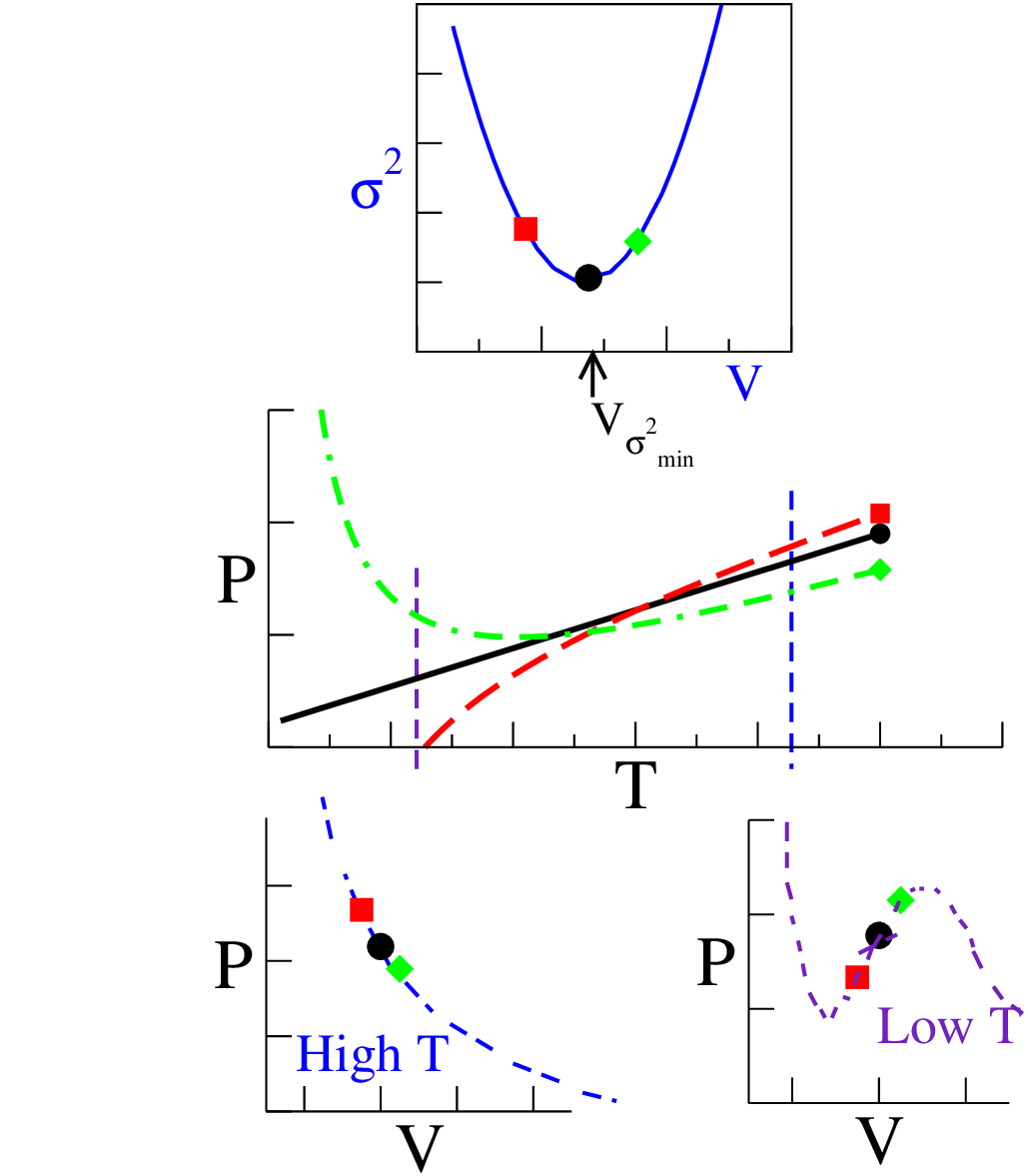}
\vspace{-0.2cm}
\caption{Schematic explanation of the connection between a minimum in $\sigma^2(V)$ and a liquid-liquid critical point. Square, circle and diamond indicate respectively a $V$ smaller than, equal to and larger than $V_{{\sigma^2}_{min}}$.  The central panel shows the corresponding isochores. Note that, as shown in Eq.\protect\ref{eq:componenti}, $P$ increases (decreases) on cooling when 
$d\sigma^2/dV>0$ ($<0$). The  bottom panels report the corresponding  high and low  $T$ isotherms. At low $T$ an unstable  Van der Waals loop appears. }
\label{fig:fig1}
\end{figure}

The harmonic gaussian landscape also elucidates
the relation between density anomalies and
the existence of a liquid-liquid critical point.
Fig.~\ref{fig:fig1} schematically compares  $P(T,V)$ along  three different isochores: at   the volume at which $\sigma^2$ is a minimum, $V=V_{\sigma^2_{min}}$, and at two other $V$-values, respectively below and above $V_{\sigma^2_{min}}$. 
In the harmonic gaussian landscape $P$
along the $V_{\sigma^2_{min}}$ isochore 
increases linearly with $T$, since, by construction,  ${\cal P}_{_{_{1/T}}}(V) =0$ and  ${\cal P}_{_{_{T}}}>0$.
For $V < V_{\sigma^2_{min}}$, $P$ decreases on cooling, since ${\cal P}_{_{_{1/T}}}(V)  <0$. The opposite trend is 
observed for  $V > V_{\sigma^2_{min}}$.
An isothermal cut of the three isochores  
shows the corresponding $P(V)$ curves.
At high $T$, where the linear $T$ term is dominant, $P(V)$  is monotonically decreasing with $V$,  while at low $T$  the $P(V)$ isotherm is not monotonic, indicating the presence of a region of instability (negative compressibility).  As a result, one must conclude that a critical point exists at an intermediate $T$,  which can be accessed if no 
other mechanism (as for example crystallization or  glass transition) preempts its observation. 

We now compare the harmonic gaussian landscape  predictions with an extensive study of the PEL 
of one of the most widely studied models for water, the 
extended simple-point charge (SPC/E)~\cite{spce}.
SPC/E  is able to reproduce water density anomalies.   The PEL for this model has been studied in some detail previously~\cite{deben,starr01}. 
We simulated a system of  216 water molecules  with classical molecular dynamics ($MD)$ in the NVE ensemble. We studied 45 different state points in supercooled states, for times longer than several $\alpha$-relaxation times. Long range forces have been modeled via the  reaction field technique. The integration time step
was 1.0 fs.  Results are averaged over 100 different independent trajectories for each state point. 
~From each of the 100 trajectories we extracted about 30 configurations and  calculated,  via conjugate gradient techniques with $10^{-15}$ tolerance,
the local minima (the so-called inherent structures) and their energies $e_{IS}$.   The vibrational density of states in the local minima has been calculated by diagonalizing the Hessian (the matrix of second derivatives of the potential). This procedure produces 3000 ($30 \times 100$) distinct  minima for each of the 45 studied state points,  allowing the statistical error to be smaller than the symbol size. The resulting  extremely accurate determination of  $\sum_{i=1}^{6N-3} \ln(\omega_i(e_{IS})) $, of the
$T$-dependence of  $\left<e_{IS}\right >$ and of the
configurational entropy $S_{conf}$~\cite{sastrynature} --- a measure of the logarithm of the number of explored PEL basins --- is crucial when accurate $V$-derivatives of the PEL parameters are required, as in the present case. 
Procedures for the evaluation of  the PEL parameters for the harmonic gaussian model have been worked out~\cite{sastrynature,eos,mossa,emipisa}. From numerical evaluation of the $T$-dependence of $\langle e_{IS}\rangle$ and $S_{conf}$  it is possible  (i) to provide evidence
that the gaussian landscape correctly describes the
SPC/E simulation data and  (ii) calculate, with a fitting procedure, $\sigma^2$ and $E_0$ and $\alpha$. 
The basin shape parameters $a$ and $b$ are calculated from the $e_{IS}$ dependence of the vibrational density of states, evaluated at the $IS$, i.e. by a linear fit of 
$\sum_{i=1}^{6N-3} \ln(\omega_i(e_{IS})) $ vs 
$e_{IS}$.
The parameters $E_0$ and $\sigma^2$ are evaluated 
via a linear fit of   $\langle e_{IS}(T) \rangle$  vs $T^{-1}$. The parameter $\alpha$ is calculated by fitting the  $T$-dependence of $S_{conf}$\cite{sastrynature,eos}.

\begin{figure}[t] 
\centering
\includegraphics[width=.30\textwidth]{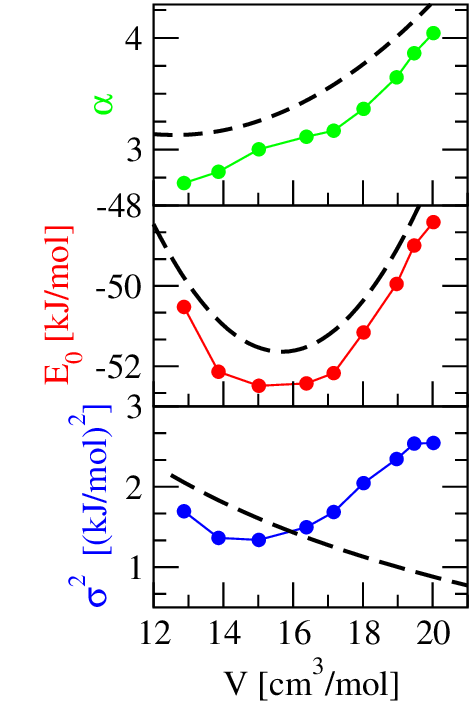}
\vspace{-0.1cm}
\caption{Volume dependence of the PEL parameters for the SPC/E potential.  Dashed lines indicates the trends expected for simple liquids\protect\cite{soft}.}
\label{fig:fig2}
\end{figure}

Fig.~\ref{fig:fig2}  shows the  $V$-dependence of the  landscape parameters  ($\alpha$, $E_0$, $\sigma^2$)
for nine different volumes and contrasts it with the
behavior characteristic of simple liquids\cite{soft}.
The total number of states $\alpha$
decreases on compressing the system, a feature common to all simple liquid models studied so far~\cite{sastrynature,eos}. 
The $V$ dependence of the energy
scale $E_0$  is also analogous to the one found in simple liquid models. Indeed $E_0$ 
first decreases on compression, corresponding to the progressive sampling of 
the most attractive part of the potential, 
then it starts to increase 
due to the sampling of the repulsive part of the potential at short intermolecular nearest neighbor distances.  
As expected on the basis of the previous discussion, the $V$-dependence of $\sigma^2$ shows instead elements which are not observed in simple liquid models, where $\sigma^2$ decreases monotonically on increasing $V$. In the SPC/E case, the variance shows a clear minimum around $V_{\sigma^2_{min}}=15$ cm$^3$/mol  (i.e. at density $\rho=1.2$ g/cm$^3$)  and hints of a  maximum at $V=20$ cm$^3$/mol ($\rho=0.9$ g/cm$^3$).  Between $15$ and $20$ cm$^3$/mol, SPC/E water exhibits density anomalies. 
The increase of $\sigma^2$ for $V>V_{\sigma^2_{min}}$
 can be  attributed to the fact that the system can  explore new configurations, characterized by the presence of
hydrogen bonds.  The formation of such bonds requires a large local volume and it is severely hampered at high density. These new accessible states widen the variance of the $\Omega(e_{IS})$ distribution, producing a
range of $V$-values where $d\sigma^2/dV$ is  positive.
When $V$ has increased to about 20 cm$^3$/mol
the system  has become mostly composed of a network of open linear hydrogen bonds and  there are no more
mechanisms available to increase the variance.

\begin{figure}[t] 
\centering
\includegraphics[width=.42\textwidth]{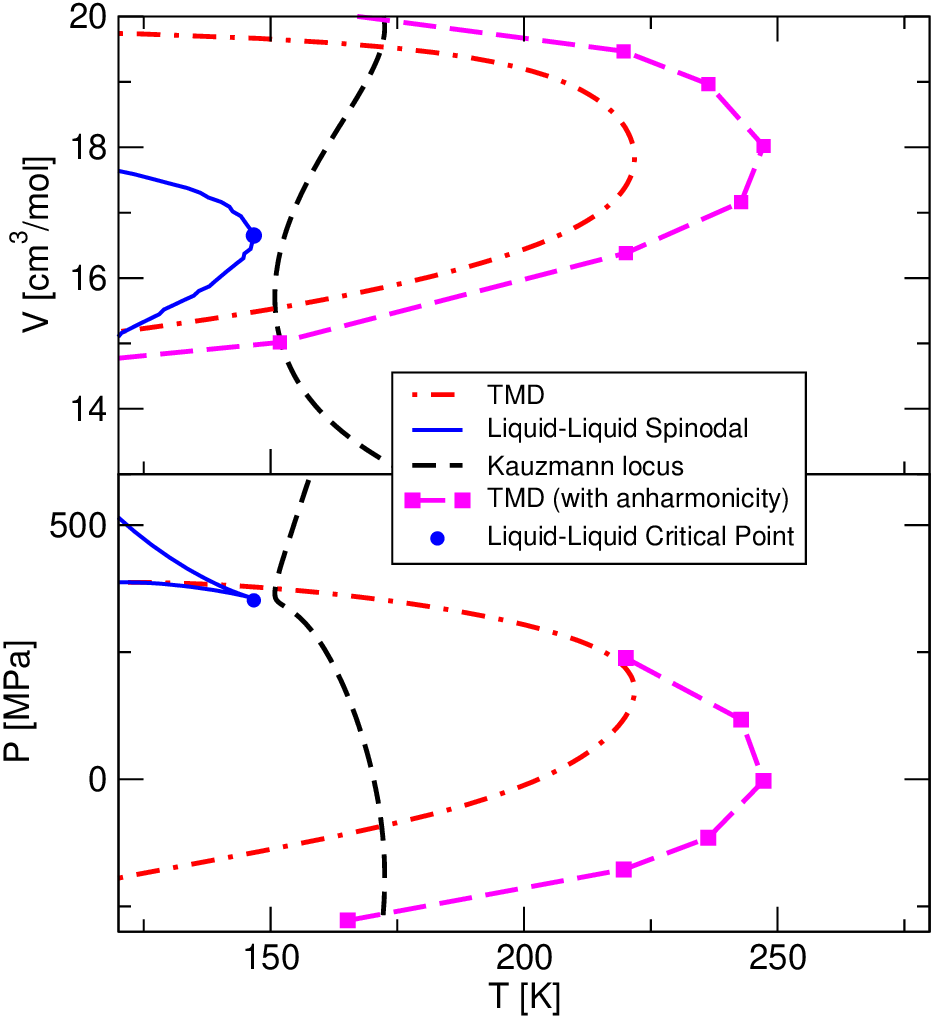}
\vspace{-0.5cm}
\caption{Harmonic gaussian landscape phase diagram for the SPC/E potential, including the $TMD$ line, the
liquid-liquid critical point with the associated spinodal
lines and the Kauzmann locus. Lines below the Kauzmann locus have no physical meaning, since they correspond to states with negative $S_{conf}$.  They are drown here to visualize the termination of the $TMD$ line and the relative position of the critical point respect to the Kauzmann locus.  Squares indicate the TMD as evaluated directly from
the MD calculations.  As discussed in the text, adding the anharmonic contributions in the theoretical calculation would move the liquid-liquid critical point into the physically meaningful $S_{conf}>0$ region.}
\label{fig:fig3}
\end{figure}

The actual  location of the critical point, which will depend on all landscape properties, can be calculated according to Eq.~\ref{eq:componenti}. For the SPC/E harmonic gaussian landscape, the resulting  phase diagram is shown in Fig.~\ref{fig:fig3}.   The density maxima locus retracing 
at low densities is consistent with the existence of
a volume beyond which the liquid returns to normal 
(i.e. with $d\sigma^2/dV<0$).   The $T$ of maximum density (TMD) locus at large pressures  terminates along the liquid-liquid spinodal, as predicted by thermodynamic consistency~\cite{speedyspinodal}.  
Fig.~\ref{fig:fig3} also show the TMD line evaluated directly from the  $MD$ $P(V,T)$ data. This curve, which properly includes  anharmonic contributions, show that  anharmonicities  do not change the topology but merely shifts the $TMD$ curve up about 25 K in $T$ and down 160 MPa in $P$.  The location of the liquid-liquid critical point for the SPC/E potential is estimated to be between $\approx 140$ and $175$ K and between $\approx 185$ and $340$ MPa, in agreement with  previous predictions~\cite{scala}.

One of the powerful features of the PEL formalism is
the possibility of simultaneously evaluating 
the EOS and $S_{conf}$.
This offers the unique possibility of evaluating,  within the chosen landscape model, the 
theoretical limit of validity of the low T extrapolations. 
In the present case, the limit  
set by the condition $S_{conf}=0$
defines a Kauzmann curve $T_K(P,V)$ below which 
the above EOS is no longer valid. 
The $T_K(P,V)$ locus is also indicated in Fig.~\ref{fig:fig3}.
The TMD locus crosses the Kauzmann locus very close 
to its reentrance, as recently predicted by Speedy~\cite{budapest}. It is interesting to observe that the location of the liquid-liquid critical point is
within the glass state and hence technically does not exist in the harmonic gaussian approximation. When anharmonic contributions are taken into account, the
critical point moves into the $S_{conf}>0$ region, suggesting
that the liquid-liquid  critical point could be a real feature of the SPC/E EOS. If this were the case, the possibility of
observing a liquid-liquid critical point in this model 
would be only hampered by the extreme slowing down of the
dynamics at low $T$ and, in principle, by crystallization\cite{tzero}.

In this Letter we have shown that, in the harmonic gaussian landscape model, density anomalies are a sufficient, but not a necessary, condition for the existence of a liquid-liquid critical point.  
Anharmonicities do not alter the picture and can be taken properly into account if necessary. Instead, the gaussian hypothesis is crucial, and models with a different distribution $\Omega(e_{IS})$  may produce density anomalies 
in systems without any liquid-liquid critical point~\cite{sastry}.   Also, high-temperature 
fluid-fluid transitions\cite{franzese} can not be described by this formalism, which is applicable only to supercooled states.    Still, the generality of the gaussian distribution, rooted in the central limit theorem strongly support the possibility that what we learn studying the gaussian landscape applies to  water.

We thank SHARCNET for a very generous allocation of
CPU time. FS thanks UWO for hospitality. We acknowledge support from MIUR Cofin 2002 and Firb and INFM Pra GenFdt.  We thank P. Poole, I. Saika-Voivod, S. Sastry and R. Speedy for discussions.


\begin{thebibliography}{99}

\bibitem{angellreview}
C. A. Angell,  in {\it Water: A Comprehensive Treatise} {\bf Vol. 7} (ed. Franks, F.) 1-81 (Plenum, New York, 1982).

\bibitem{poole}
P. H. Poole  {\it et al.},
{ Nature} {\bf 360}, 324 (1992).

\bibitem{ponyatonski} 
E.  G. Ponyatovsky,  V. V.   Sinand and  T. A. Pozdnyakova, 
{ JEPT Lett.} {\bf 60}, 360  (1994).

\bibitem{mix}
P. H. Poole {\it et al.}, Phys. Rev. Lett. {\bf 73} , 1632 (1994);
S. Sastry {\it et al.},  Phys. Rev. E {\bf 53}, 6144 (1996); E. La Nave {\it et al.}, Phys. Rev. E{\bf 59},
6348 (1999); G. Franzese {\it et al.}, Phys. Rev. E {\bf 67}, 011103 (2003).
 
\bibitem{mishimastanley} 
O.  Mishima  and  H. E.  Stanley
 { Nature} {\bf 392}, 164 (1998).


\bibitem{poolepre} P. H. Poole {\it et al.}
{ Phys. Rev. E } {\bf 48},  4605  (1993).

\bibitem{mishima} 
 O.  Mishima, L. D.   Calvert   and   E. Whalley,  
{ Nature} {\bf 314},  76  (1985).

\bibitem{sw}
F. H. Stillinger and  T. A. Weber, 
{ Science} {\bf 225}, 983  (1984).


\bibitem{speedy}
R. J.  Speedy,
{  J. Phys. Chem. B}  {\bf 105},  11737  (2001).

\bibitem{stillingerjpc}
 F. H. Stillinger, 
{ J. Phys. Chem. B} {\bf 102}, 2807 (1998).

\bibitem{lewis}
 P. G. Debenedetti {\it et al}
{ Adv. Chem. Eng.} {\bf 28},  21  (2001).

\bibitem{sastrydeben}
S. Sastry,  P. G.  Debenedetti  and F. H. Stillinger, 
{ Nature} {\bf 393},  554 (1998).

\bibitem{ivan}
I. Saika-Voivod,  P. H.  Poole   and  F. ~Sciortino, 
{ Nature}  {\bf 412},  514 (2001).

\bibitem{mossa}  S. Mossa {\it et al.}
{ Phys. Rev. E} {\bf 65},  041205  (2002).


\bibitem{sastrynature}
S. Sastry, 
{  Nature}  {\bf 409}, 164 (2001).

\bibitem{starr01}
F. W.  Starr  {\it et al}
{ Phys. Rev. E} {\bf 63}, 041201 (2001).
%
 
\bibitem{rem} 
B.  Derrida,  
{ Phys. Rev. B} {\bf 24},  2613  (1981).

\bibitem{sasai} 
M. Sasai,   J. Chem. Phys. {\bf 118}, 10651 (2003).



\bibitem{heuer00}
A. Heuer  and  S. Buchner,   
{ J. Phys. Cond. Matter} {\bf 12}, 6535  (2000).



\bibitem{keyes}
T. Keyes,    
{ Phys. Rev. E} {\bf 62}, 7905 (2000).



\bibitem{emipisa}
 E. la Nave {\it et al}, 
J. Phys: Condens. Matter {\bf 15}  1085 (2003).



\bibitem{eos} 
E. La Nave {\it et al }
{ Phys. Rev. Lett.} {\bf 88},  225701[1] (2002).


\bibitem{nota1}
The possibility of monotonic decrease of $P$ with $T$ or of a maximum must be rejected since it would not reproduce the correct high $T$ behavior  (in other words, $ {\cal P}_{_{_{T}}}$ must be positive).

\bibitem{nota2}
We note that  the structure of Eq.~\ref{eq:componenti}  suggests that density minima  (maxima in isochoric $P(T)$)  can not be present.  We also remind the possibility
of creating density anomalies  from vibrational properties.
Indeed, in some crystals anharmonicities in the vibration may produce weak density anomalies. 



\bibitem{spce}

H. J. C. Berendsen ,  J. R. Grigera and T. P. Straatsma, 
{ J. Phys. Chem.} {\bf 91}, 6269  (1987). 
 

\bibitem{deben}
P. G. Debenedetti {\it et al}
{ J. Phys. Chem. B}  {\bf 103}, 7390 (1999).


\bibitem{soft}
Within the gaussian harmonic model, it is possible to work out analitically the $V$-dependence of the statistical properties of the landscape in the case of soft-sphere potentials ($V(r) \sim r^{-n}$).  $\alpha$ is found to be constant, $E_0 \sim V^{-n/3}$ and $\sigma^2  \sim V^{-2n/3}$\cite{scott}. For Lennard Jones systems\cite{sastrynature,eos}, it has been found numerically that $\alpha$ increases with $V$, 
$\sigma^2$ decreases and $E_0$ displays a minimum.

\bibitem{speedyspinodal}
R. J. Speedy, 
{ J. Phys. Chem. } {\bf 86}, 982 (1982). 

\bibitem{scala}
A.  Scala, {\it et al.} 
 { Phys. Rev. E }  {\bf 62}, 8016 (2000).


\bibitem{budapest} 

R. J. Speedy, 
in { \it Liquids Under Negative Pressure}. 
(eds. Imre, A. R., Maris, H.J. and Williams, P.R. ), 1-12
(Kluwer Academic Publisher,  Boston, 2002 ).
\bibitem{tzero}
At low temperature, basins in the tail of $\Omega(e_{IS})$ distributions are explored and hence the validity of the gaussian approximation is questionable.  $T_K$ could be indeed located at much lower $T$, if not at $T=0$.

\bibitem{sastry}  
 S. Sastry {\it et al}
 Phys. Rev. E {\bf 53}, 6144 (1996).

\bibitem{franzese}

G. Franzese {\it et al.},  Nature  {\bf 409}, 692 (2001).

\bibitem{scott} 
M. S. Shell et al, 
J. Chem . Phys.  {\bf 118} , 8821 (2003).

\end{thebibliography}
 \end{document}